\documentclass[twocolumn,showpacs,amsmath,amssymb,pra]{revtex4}
\usepackage[latin9]{inputenc}
\usepackage{textcomp}
\usepackage{amsmath}
\usepackage{amssymb}
\usepackage{graphicx}
\usepackage{epstopdf}

\usepackage{bbm}
\usepackage[toc,page,title,titletoc,header]{appendix}
\usepackage{dcolumn}\usepackage{bm}
\newcommand{\be}{\begin{equation}}
\newcommand{\ee}{\end{equation}}
\newcommand{\bey}{\begin{eqnarray}}
\newcommand{\eey}{\end{eqnarray}}
\newcommand{\bw}{\begin{widetext}}
\newcommand{\ew}{\end{widetext}}
\newcommand{\ww}{\widetilde}

\newcommand{\ov}{\overline}

\newcommand{\ra}{\rangle}
\newcommand{\la}{\langle}

\newcommand{\C}{{\cal{C}}}

\newcommand{\ba}{\begin{array}}
\newcommand{\ea}{\end{array}}
\newcommand{\bi}{\begin{itemize}}
\newcommand{\ei}{\end{itemize}}
\newcommand{\bem}{\begin{enumerate}}
\newcommand{\eem}{\end{enumerate}}

\makeatother

\begin{document}

  \title{Correlations in eigenfunctions of quantum chaotic systems with sparse Hamiltonian matrices
 }

 \author{Jiaozi Wang and Wen-ge Wang
 \footnote{ Email address: wgwang@ustc.edu.cn}
  }

\affiliation{Department of Modern Physics, University of Science and Technology of China,
Hefei, 230026, China}

\date{\today}

\begin{abstract}
  In most realistic models for quantum chaotic systems, the
  Hamiltonian matrices in unperturbed bases have a sparse structure.
  We study correlations in eigenfunctions of such systems and
 derive explicit expressions for some of the correlation functions with respect to energy.
  The analytical results are tested in several models by numerical simulations.
  An application is given for a relation between transition probabilities.
\end{abstract}


\maketitle


 \section{Introduction}

 Statistical properties of energy eigenfunctions (EFs) of quantum chaotic systems
 have been studied extensively in the past years
 \cite{Berry77,Haake,CC94book,Meredith98,Buch82,Sr96,Iz96,Connor87,Sr98,Backer02,Urb03,
 scanz05,Mirlin00,Mirlin02,KpHl,Kp05,Bies01,Falko96,
 Prosen03,Lewf95,Prig95,Wg98,Gnutzmann10,Anh11,stockmann09,Kaplan09}.
 They are of interest in various fields of physics and have many applications, e.g., in
 statistical and transport properties in chaotic quantum dots \cite{qd1,qd2},
 in wave functions in optical, elastomechanical,
 and microwave resonators \cite{opt1,opt3,mech1,micw2,micw3,micw4,ccchen},
 and in the decay and fluctuations of heavy nuclei \cite{nuc1,nuc2,nuc3}.
 In particular, they play an important role in the understanding of thermalization
 \cite{Deutsch91,Srednicki94,Rigol08,Haake12,Izrailev12}.

 Due to the remarkable success of the random-matrix theory (RMT) in the description of statistical
 properties of energy levels of quantum chaotic systems
 \cite{Haake,CC94book,Mirlin00,Haake-RMT}, it would be
 natural to expect that the RMT may be useful in
 the description of statistical properties of EFs in these systems.
 Indeed, this expectation has led to some successful applications
 (see, e.g., reviews given in Refs.\cite{Mirlin00,Gorin-rep}).
 In fact, restricted to main bodies of EFs \cite{Buch82}, or to the so-called nonperturbative
 regions of the EFs \cite{wwg-LMG02,chaosdis}, numerical simulations show
 that the distribution of the components of the EFs has a Gaussian shape, as predicted by the RMT.
 But, deviation from the Gaussian distribution has been observed,
 when the tail regions of EFs are taken into account \cite{Meredith98}.

 Consistently,  for EFs in the configuration space, Berry's conjecture assumes uncorrelated phases
 for their components in the momentum representation \cite{Berry77}.
 Based on Berry's conjecture and semiclassical analyses,
 it has been found that neighboring EFs in many-body systems predict
 similar results for local observables \cite{Srednicki94}.
 This property, which has also been found in a RMT study \cite{Deutsch91},
 is of relevance to thermalization
 and, in a broader situation, is nowadays referred to as eigenstates thermalization hypothesis (ETH)
 \cite{Rigol08}.
 Furthermore, when specific dynamics, e.g., periodic orbits and long-range correlations are taken into account,
 modifications should be introduced to Berry's conjecture
 \cite{KpHl,Bies01,Sr98,Backer02,Urb03}.

 In fact, for EFs in chaotic many-body quantum systems,
 correlations more than that predicted by the original RMT have been found and
 modified versions of the RMT have been investigated
 \cite{EGOE71,EGOE03,EGOE10,GGW98}.
 For example, contrary to the vanishing correlation function predicted by the RMT,
 in a many-body system with a sparse Hamiltonian matrix,
 non-vanishing four-point correlations have been observed,
 which are of relevance to important physical quantities such as transition probabilities  \cite{Iz96}.
 Moreover, correlations have been studied for operators at different times in
 a two dimensional kicked quantum Ising model \cite{Prosen14}.

 In this paper, we study correlations among components of EFs,
 particularly the phase correlations,  in quantum chaotic systems
 whose Hamiltonian matrices have a sparse structure in unperturbed bases.
 Such a sparse structure is commonplace in realistic models.
 Under this structure, each unperturbed state is coupled to
 a small fraction of other unperturbed states.
 As a result, it is reasonable to expect certain correlations among components of the EFs,
 as shown in the example mentioned above in  Ref.\cite{Iz96}.
 We'll derive explicit expressions for some of the correlation functions
 and test the results by numerical simulations.
 We also discuss an application of the results.

 The paper is organized as follows.
 In Sec.{\ref{sect-II}},  we discuss the models to be employed.
 Section \ref{sect-predis} is devoted to generic discussions for the
 type of correlation functions to be studied.
 Then, some specific correlation functions are discussed in Sec.{\ref{sect-cf-samesign}},
 for the case in which the perturbation matrix has elements with a homogeneous sign.
 The case with nonhomogeneous signs of the matrix elements is discussed in Sec.{\ref{sect-cf-randomsign}}.
 An application is given in Sec.{\ref{sect-apply}} for a relation between some transition probabilities.
 Finally, conclusions are given in Sec. {\ref{sect-sum}}.

\section{Models employed}\label{sect-II}

 We consider quantum chaotic systems, for each of which the Hamiltonian is written as $H=H_0+ V$,
 where $H_0$ is an unperturbed Hamiltonian and $V$ indicates a perturbation.
 We'll employ four models in our numerical simulations.
 Parameters in the four models are set, such that they are in the quantum chaotic regime,
 in which the distribution of the nearest-level spacings is close to the prediction of the RMT.

The first model we consider is a three-orbital LMG model\cite{LMG}.
 This model is composed of $\Omega$ particles, occupying three energy levels labeled by $r=0,1,2$, each with
 $\Omega$-degeneracy.
 Here, we are interested in the collective motion of this model.
 We use $\epsilon_{r}$ to denote the energy of the $r$-th level
 and, for brevity, we set $\epsilon_{0}=0$.
 The Hamiltonian of the model is written as
\begin{equation}
H=H_{0}+ V,
\end{equation}
 where $H_0$ and $V$ are the unperturbed Hamiltonian and the perturbation, respectively,
\begin{equation}
 H_{0}=\epsilon_{1}K_{11}+\epsilon_{2}K_{22}, \quad V=\sum_{t=1}^{4}\mu_{t}V^{(t)}.
\end{equation}
 Here, $K_{rr} $ represents the particle number operator for the level $r$ and
\begin{eqnarray}
V^{(1)}=K_{10}K_{10}+K_{01}K_{01},\ V^{(2)}=K_{20}K_{20}+K_{02}K_{02},\nonumber \\
 V^{(3)}=K_{21}K_{20}+K_{02}K_{12},\ V^{(4)}=K_{12}K_{10}+K_{01}K_{21}, \ \
\end{eqnarray}
 where $K_{rs}$ with $r\neq s$ indicate particle raising and lowering operators.
 In our numerical simulations, the particle number is set $\Omega=40$,
 as a result, the Hilbert space has a dimension $861$.
 Other parameters are $\epsilon_{1}= 1.10, \epsilon_{2} = 1.61,
\mu_{1} = 0.031, \mu_{2} = 0.035, \mu_{3} = 0.038$, and $\mu_{4} = 0.033$.
 In the computation of the correlation functions,
 averages were taken over $50$ perturbed eigenstates $|E_\alpha\ra$ in the middle energy region.


 The second model is a single-mode Dicke model\cite{Dicke,Emary2003}, which
 describes the interaction between a single bosonic mode and a collection of $N$ two-level atoms.
 The system can be described in terms of the collective operator ${\bf \hat{J}}$ for the $N$
 atoms, with
 \be
 \hat{J}_{z} \equiv \sum_{i=1}^{N} \hat{s}_{z}^{(i)},\ \ \hat{J}_{\pm} \equiv \sum_{i=1}^{N}
 \hat{s}_{\pm}^{(i)},
 \ee
 where $\hat{s}_{x (y,z)}^{(i)}$ are Pauli matrices divided by $2$ for the $i$-th atom.
 The Dicke Hamiltonian is written as~\cite{Emary2003}
\be
H=\omega_0 J_z +\omega a^{\dagger}a+\frac{\lambda}{\sqrt{N}}(a^{\dagger}+a)(J_+ +J_-).
\ee
 In the resonance condition, $\omega_0 =\omega$.
 The operators $J$ obey the usual commutation rules for the angular momentum,
\be
[J_z , J_{\pm}]=\pm J_{\pm},\ \ [J_+,J_-]=2J_z.
\ee
 We write the Hamiltonian in the form $H = H_0+ V $, with
 $H_0  =  \omega_0 J_z +\omega a^{\dagger}a $.
 In numerical simulations, we take $N=40$ and $\lambda=1$, and
 the particle number of the bosonic field is truncated at $n=40$.

 The third model is a modified XXZ model, called a defect XXZ model\cite{DX},
 in which two additional magnetic fields are applied to two of the $N$ spins in the XXZ model,
\begin{gather}\label{} \nonumber
H= \mu_1 \sigma^1_z+ \mu_{4} \sigma^{4}_z
 \\ + \sum_{i=1} ^{N-1} {[J ( \sigma^i _x \sigma^{i+1} _x
+ \sigma^i _y \sigma^{i+1} _y )+ \mu \sigma^i _z \sigma^{i+1} _z ]}.
\end{gather}
 Without the additional magnetic fields, the system is integrable.
 We also write $H=H_0 + V$, where
\bey
H_0  =  \mu_1 \sigma^1_z+ \mu_{4} \sigma^{4}_z+\sum_{i=1}^{N-1}\mu \sigma^i_z
\sigma^{i+1}_z. \
\eey
 The total Hamiltonian $H$ is commutable with $S_z$, the $z$-component of the total spin,
 and we use the subspace with $S_z = -2$ in our numerical study.
 Parameters used in this model are $N=12$, $\mu_1 = \mu_4 =1.11$, $\mu = 0.5$,
 and $J =1.4$.

The last model we employ is a modified 1D-Ising chain in transverse field, called a defect
 Ising model\cite{DI}, with the Hamiltonian
\be
H  =  \mu_1 \sigma^1_z+ \mu_{4} \sigma^{4}_z+\sum^N_i J_z \sigma^i_z
\sigma^{i+1}_z+ \lambda \sum_{i=1}^{N-1} { \sigma^i _x }.
\ee
 In the form of $H=H_0 + V$,
\bey
H_0 & = & \mu_1 \sigma^1_z+ \mu_{4} \sigma^{4}_z+\sum^N_i J_z \sigma^i_z \sigma^{i+1}_z.
\eey
 Parameters used in this model are $N=10$, $\mu_1 = \mu_4 = 1.11$, $J_z=1$, $\mu_z=0.3$,
 and $\lambda = 0.45$.

\section{Generic discussions about correlation function}\label{sect-predis}

 In this section, we discuss the type of correlation function to be studied in this paper.
 We use $|E_\alpha\ra$ and $|E^0_i\ra$ to denote eigenstates of
 the perturbed Hamiltonian $H$ and of the unperturbed one $H_0$, respectively,
\begin{gather}\label{}
 H|E_\alpha\ra = E_\alpha|E_\alpha\ra, \quad H_0|E^0_i\ra = E^0_i|E^0_i\ra,
\end{gather}
 with eigenenergies in increasing order,
 and use $V_{ij}$ to denote elements of the perturbation, $V_{ij}=\la E^0_i|V|E^0_j\ra$.
 We assume that the perturbation $V$ has a sparse structure in the eigenbasis of $H_0$,
 that is, $V_{ij}=0$ for most of the pairs $(i,j)$.
 (All the four models discussed in the previous section have this property.)
 We also assume that the perturbation has vanishing
 diagonal elements, $V_{ii}=0$ \cite{foot-Vii}.
 The expansion of $|E_\alpha\ra$ in $|E^0_i\ra$ is written as
\begin{gather}\label{}
 |E_\alpha\ra = \sum_i C_{\alpha i}|E^0_i\ra,
\end{gather}
 where the components $C_{\alpha i} = \la E^0_i|E_\alpha\ra$ give the EF.
 For the sake of simplicity in discussion, we assume that the system has the time-reversal symmetry
 and the elements $V_{ij}$, as well as the components $C_{\alpha i}$ are real.

 Physically, the following transition amplitude is of interest,
\begin{gather}\label{Fij}
 F_{ij}(t)= \la E^0_j|U(t)|E^0_i\ra,
\end{gather}
 where $U(t) = e^{-iHt}$.
 Straightforward derivation shows that
\begin{gather}
 F_{ij}(t) = \sum_\alpha e^{-iE_\alpha t} C_{\alpha j} C_{\alpha i}.
\end{gather}
 When the time $t$ is not long, neighboring levels $E_\alpha$ give similar contributions
 to the phase of $e^{-iE_\alpha t}$.
 Suppose that the average of $C_{\alpha j} C_{\alpha i}$ over neighboring levels
 can be approximately treated as a smooth function of the energy $E_\alpha$, denoted by $\C (E_\alpha)$,
 which approximately holds  for most EFs in quantum chaotic systems.
 Then,
\begin{gather*}
 F_{ij}(t) \approx \sum_\alpha e^{-iE_\alpha t} \C(E_\alpha)
 \to \int dE e^{-iE t} \C(E) \rho(E),
\end{gather*}
 where $\rho(E)$ indicates the density of states.
 Therefore, knowledge about the function $\C(E)$, which is in fact a correlation function,
 is useful in the study of physical quantities such as transition probabilities.

\begin{figure}
\includegraphics[width=\columnwidth]{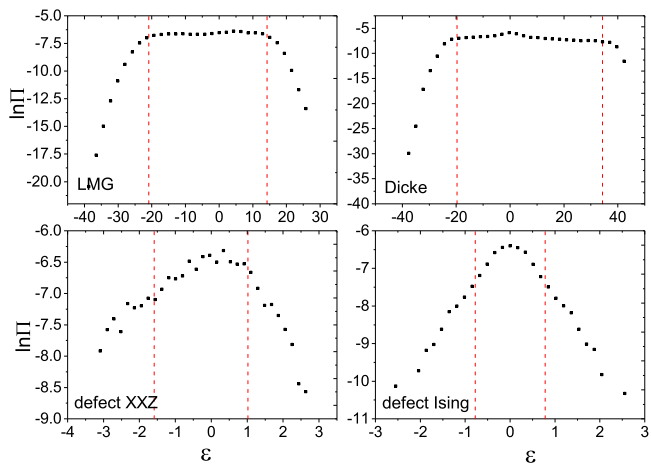}
\caption{ (Color online) Averaged shape of EFs in the four models as functions of the
energy difference $\varepsilon$, in the logarithmic scale.
 The average was taken over 50 EFs in the middle energy region in each model.
  Vertical straight lines indicate the edges of the (smallest) non-perturbative regions
  discussed in Refs.\cite{Wg98,wwg-GBWPT}.
 }\label{EF-shape}
\end{figure}

 For the above-discussed reason, we
 study correlation functions as an average of $C_{\alpha j} C_{\alpha i}$  with respect to the energy.
 It is known that, usually, the EF of $|E_\alpha\ra$ is approximately centered at $E_\alpha$
 (see Fig.\ref{EF-shape} for examples of the averaged shapes of EFs).
 Therefore, it is convenient to consider correlations as functions of the energy difference
 between perturbed and unperturbed states, namely,
 as functions of $\varepsilon_{\alpha l} \equiv E_{l}^{0} - E_{\alpha}$.
 The average, which is used in the computation of the correlation functions,
 is taken over the perturbed energy $E_\alpha$ for a fixed value of $\varepsilon$.
 Determination of the label $l$ will be specified below,
 when discussing specific correlation functions.

 We find that correlation functions  behave differently for labels $i$ and $j$ coupled in different ways.
 Therefore, we study correlation functions according to the ways of coupling.
 Specifically, we use $S_n$ to denote the set of those pairs $(i,j)$,
 for each of which the two unperturbed states
 $|E_i^0\ra$ and $|E_j^0\ra$ have an ``$n$-step'' coupling, that is,
\begin{equation}\label{}
 S_n = \{ (i,j): (V^n)_{ij} \ne 0, \ (V^m)_{ij} = 0 \ \text{ for} \ 0<m < n \ \}.
\end{equation}
 We call a correlation function, which is computed for pairs $(i,j)$ belonging to
 a given set $S_n$, an \emph{$n$th-order} correlation function.

 For example, the {first-order correlation function} is defined by
\begin{equation}\label{C1}
 {\cal C}_1(\varepsilon) = \la C_{\alpha i}C_{\alpha j}\ra / \Pi(\varepsilon)
 \quad \text{for} \ (i,j)\in S_1,
\end{equation}
 where $\Pi(\varepsilon) $ indicates the averaged shape of the EFs,
 $\Pi(\varepsilon)= \la |C_{\alpha i}|^2 \ra$.
 Here and hereafter,
 for an average indicated by $\la \ \ra$, we take $|E^0_l\ra = |E^0_i\ra$
 for $\varepsilon_{\alpha l} $ discussed above.
 The second-order correlation function is defined by
\be \label{C2}
 {\cal C}_{2}(\varepsilon) = \la C_{\alpha i}C_{\alpha j}\ra' /{\Pi(\varepsilon)}
 \quad \text{for} \ (i,j)\in S_2,
\ee
 where the prime in $\la \ \ra'$ indicates an average  for which the
 labels $l$  in $\varepsilon_{\alpha l} $ satisfy $V_{il}V_{lj} \ne 0$.

\section{Correlation functions for $V_{ij}$ with homogeneous sign}\label{sect-cf-samesign}

 In this section, we discuss correlation functions for perturbations $V$,
 whose nonzero elements have a homogeneous sign.

 \subsection{The first-order correlation function}

\begin{figure}
\includegraphics[width=\columnwidth]{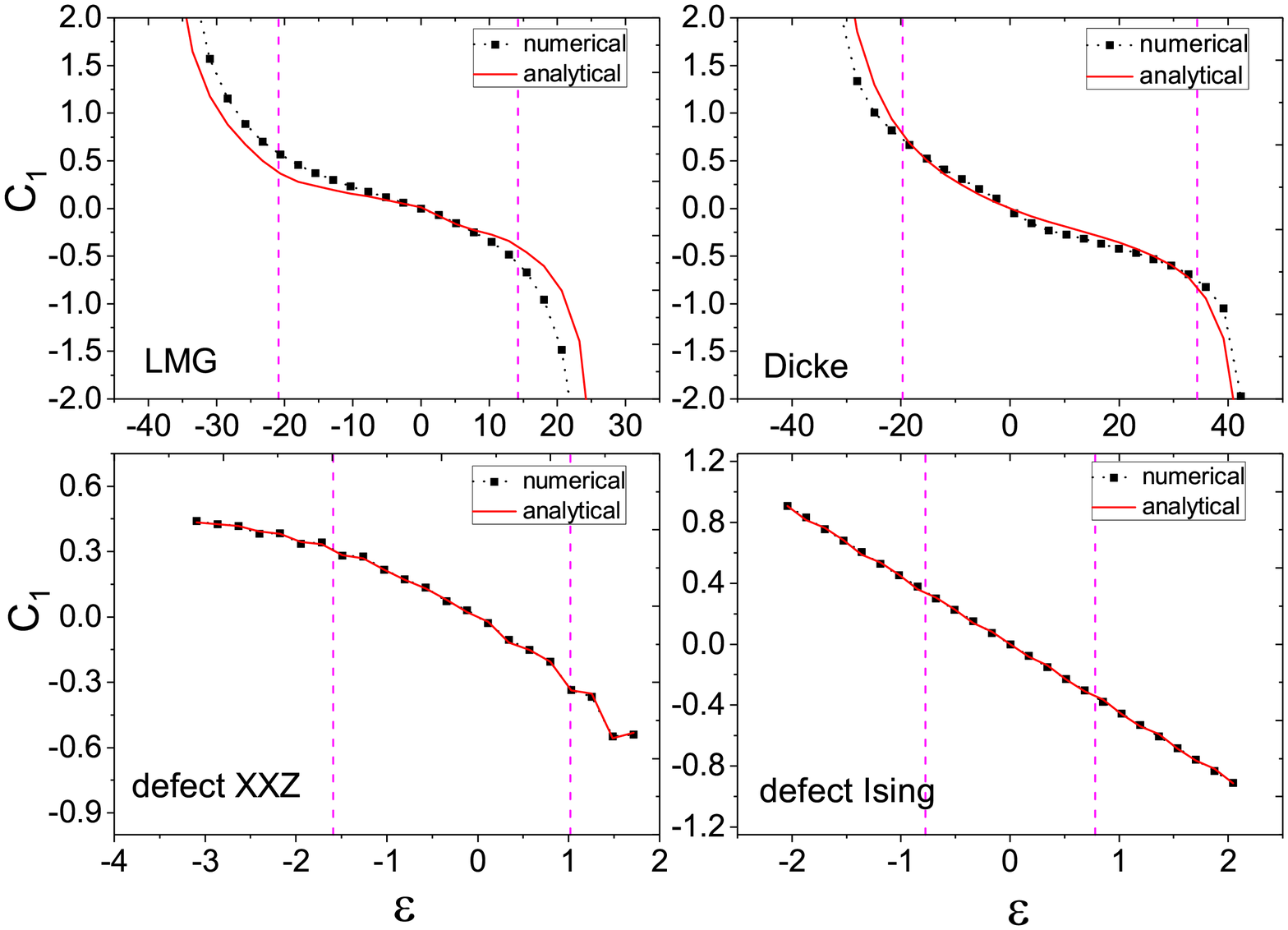}
\caption{ (Color online) The first-order correlation function ${\cal C}_1$ in Eq.(\ref{C1})
  for $(i,j) \in S_1$,
  as a function of the energy difference $\varepsilon$, in four models.
  Solid curves indicate predictions of Eq.(\ref{C1-expr}).
  (See Fig.\ref{EF-shape} for the meaning of the vertical dashed lines.)
  }\label{fig-C1}
\end{figure}

 To find an expression for the correlation function ${\cal C}_1(\varepsilon)$,
 let us write the stationary Schr\"{o}dinger equation, $H|E_\alpha\ra = E_\alpha|E_\alpha\ra$, in the form,
\be \label{Seq}
 C_{\alpha i}= -\frac{1}{\varepsilon_{\alpha i}} \sum_{j\in g_i}V_{ij}C_{\alpha j},
\ee
 where $g_i$ indicates the set of those labels $j$ for which $V_{ij}$ are nonzero,
 namely,
\begin{gather}\label{}
  g_i = \{ j:V_{ij}\neq 0 \}.
\end{gather}

 Multiplying both sides of Eq.(\ref{Seq}) by $C_{\alpha i}$, then, taking the average $\la \ \ra$
 one gets
\begin{eqnarray}\label{C2-imm}
 \la |C_{\alpha i}|^2 \ra = -\frac{1}{\varepsilon} \ov N
 \la V_{ij}C_{\alpha i}C_{\alpha j} \ra,
\end{eqnarray}
 where $\ov N = \la\sum_{j\in g_i} 1 \ra$
 is the average number of coupling to one unperturbed state.
 For quantum chaotic systems,  when the fluctuations of nonzero $V_{ij}$ are not very strong,
 the average over $V_{ij}C_{\alpha i}C_{\alpha j}$ can be taken separately for $V_{ij}$
 and $C_{\alpha i}C_{\alpha j}$, giving
\begin{equation}\label{VCC}
 \la V_{ij}C_{\alpha i}C_{\alpha j}\ra \simeq \ov V \la C_{\alpha i}C_{\alpha j}\ra,
\end{equation}
 where $\ov V = \langle V_{ij} \rangle$ for $V_{ij} \ne 0$.
 Then, Eq.(\ref{C2-imm}) gives
 \begin{equation}\label{C1-expr}
 {\cal{C}}_{1}(\varepsilon) \simeq - \frac{\varepsilon}{\ov V \ \ov N}.
\end{equation}

 An interesting feature can be seen from Eq.(\ref{C1-expr}),
 that is, in the case that $\ov V$ and $\ov N$ change slowly with $\varepsilon$,
 the first-order correlation function ${\cal C}_1$ is almost linear in $\varepsilon$.
 Thus, at $\varepsilon$ close to $0$,
 the two components $C_{\alpha i}$ and $C_{\alpha j}$ have almost uncorrelated signs,
 while, for $|\varepsilon|$ not small, $|{\cal C}_1|$ can be obviously larger than zero
 and $C_{\alpha i}C_{\alpha j}$ tend to have the same sign as $(-\varepsilon \text{sign}(V_{ij}))$.

\begin{figure}
\includegraphics[width=\columnwidth]{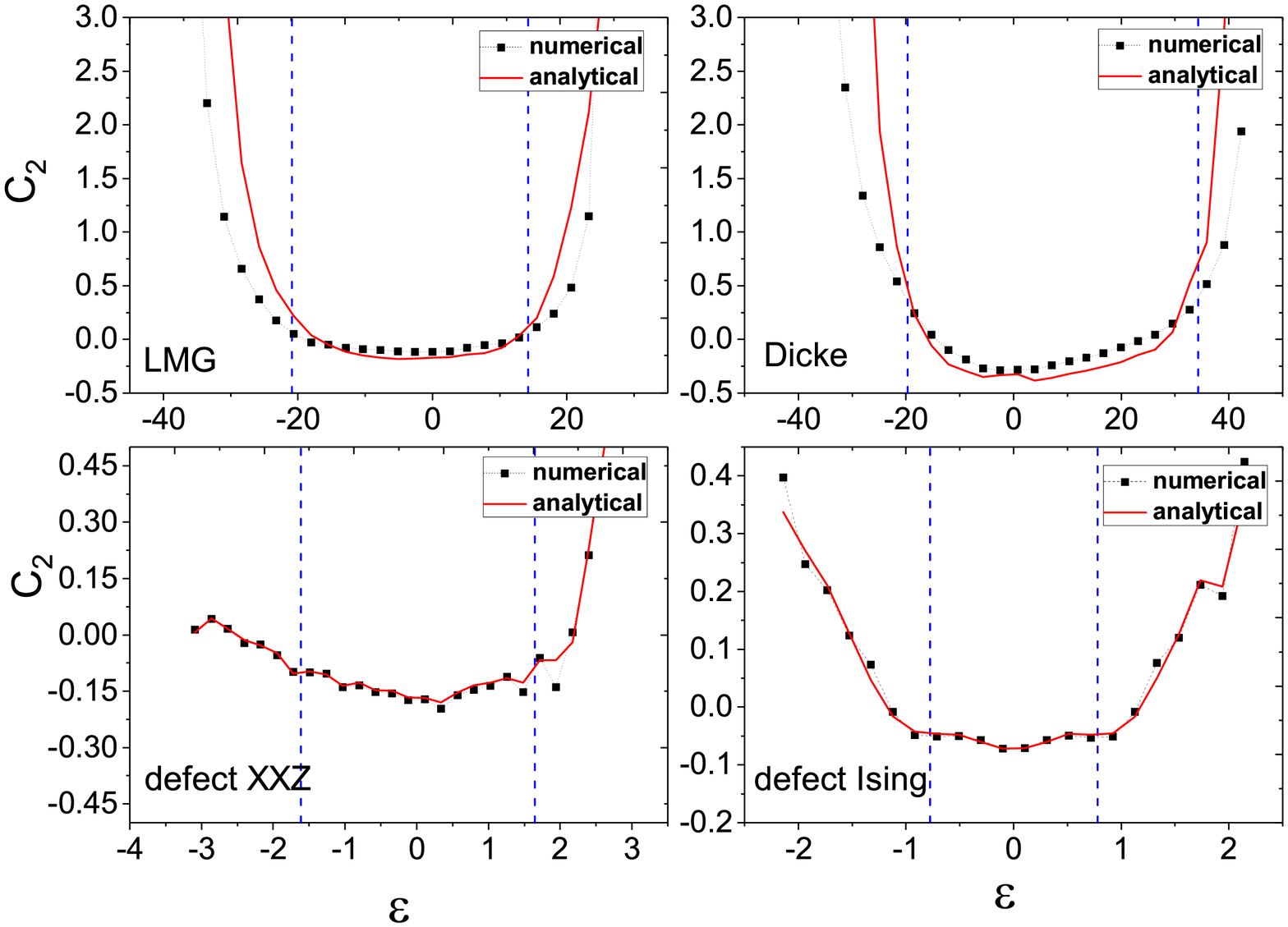}
\caption{ (Color online)
Similar to Fig.\ref{fig-C1}, but for the correlation function ${\cal C}_2$
in Eq.(\ref{C2}) with predictions given by Eq.(\ref{C2-expr}).
 }\label{fig-C2}
\end{figure}

 Numerical simulations have been performed in the four models discussed previously
 to check the predictions given above.
 In all the four models, nonzero $V_{ij}$ have the positive sign.
 In the two defect spin models, nonzero elements $V_{ij}$ share the same value,
 and good agreement between direct numerical simulations and the prediction
 of Eq.(\ref{C1-expr}) has been observed
 in the whole regime of $\varepsilon$ (Fig.\ref{fig-C1}).

 In the two models of LMG and Dicke, nonzero elements $V_{ij}$ have fluctuations,
 being stronger in the LMG model.
 In these two models, the agreement between numerical simulations and analytical predictions
 is good in the central region of the EFs, but, is not so good in the long-tail regions
 with large $|\varepsilon|$.
 For comparison, we have also computed the correlation function for the set composed
 of all the pairs $(i,j)$ and found that it is close
 to zero as predicted by the RMT, except in the long-tail regions of the EFs
 in which a perturbation theory is valid \cite{Wg98}.

 The averaged shape of the EFs, namely, $\Pi(\varepsilon)$, are plotted in Fig.\ref{EF-shape}.
 In both the LMG and the Dicke models, $\Pi(\varepsilon)$ has a platform in the central
 region, with long tails decaying exponentially.
 While, it is approximately exponentially-localized in the defect Ising model and
 is partially so in the defect XXZ model.
 This difference is related to the fact that the Hamiltonian matrices
 in the two former models have a clear band structure, but those in the two latter models do not.
 In all the four models, main bodies of the EFs lie within the so-called non-perturbative regions
 predicted by a generalized Brillouin-Wigner perturbation theory \cite{Wg98,wwg-LMG02,wwg-GBWPT},
 which are indicated by vertical dashed lines in the figures.
 Making use of components inside the non-perturbative region of an EF,
 components outside it can be expanded in a convergent perturbative expansion.

 \subsection{The second-order correlation function}

 To find an expression for the second-order correlation function,
 let us consider a label $k$, for which $V_{ik}V_{kj}\ne 0$.
  Making use of Eq.(\ref{Seq}), one gets
\begin{equation}\label{CC2}
|C_{\alpha k}|^{2}  =  \frac{1}{\varepsilon^{2}_{\alpha k}}\sum_{i}|V_{ki}|^{2}|C_{\alpha i}|^{2}
  +\frac{1}{\varepsilon^{2}_{\alpha k}}\sum_{i\neq j}V_{ki}V_{kj}C_{\alpha i}C_{\alpha j}. \ \
\end{equation}
 Taking the average $\la \ \ra'$ on both sides of Eq.(\ref{CC2})
 and following arguments similar to those leading to Eq.(\ref{C1-expr}), one gets
\be \label{C2-der1}
\Pi(\varepsilon) \simeq \frac{\ov{V^2}}{\varepsilon^2 }\Pi_d(\varepsilon) \ov N
+\frac{\ov{W}}{\varepsilon^2 }\ov N (\ov N -1) \la
C_{\alpha i}C_{\alpha j}\ra',
\ee
 where $\ov{V^2} = \la V_{ij}^{2}\ra$,
 $\ov W = \la V_{ki}V_{kj} \ra'$,
 and $\Pi_d(\varepsilon) \equiv \la |C_{\alpha i}|^{2}\ra'$.
 Note that $\Pi_d(\varepsilon)$ is not exactly the same as $\Pi(\varepsilon)$.

 Writing $\Pi_{d}(\varepsilon) = \eta \Pi(\varepsilon)$,
 we get the following expression of ${\cal C}_2$,
\be \label{C2-expr}
 {\cal C}_{2}(\varepsilon) \simeq \frac{\varepsilon^2 - \ov{V^2} \ov N \eta}
 {\ov W \ \ov N  (\ov N-1)},
\ee
 showing a quadratic dependence on $\varepsilon$.
 According to Eq.(\ref{C2-expr}), the two components
 $C_{\alpha i}$ and $C_{\alpha j}$ of $(i,j) \in S_2$ have a sign correlation different from
 that for $(i,j)$ in the set $S_1$ discussed above.
 For example, for $\varepsilon$ around $0$, the average of $C_{\alpha i} C_{\alpha j}$ of $(i,j) \in S_2$
 has a minus sign.

 Numerical tests for the prediction in Eq.(\ref{C2-expr}) are shown in Fig.\ref{fig-C2}.
 Similar to the case of first-order correlation discussed above,
 in the two defect spin models, good agreement has been observed
 in the whole regime of $\varepsilon$.
 In the two models of LMG and Dicke, the agreement
 is relatively good in the central region of the EFs, but, is not good in the long-tail regions
 with large $|\varepsilon|$ where a perturbative treatment is valid.

\begin{figure}
\includegraphics[width=\columnwidth]{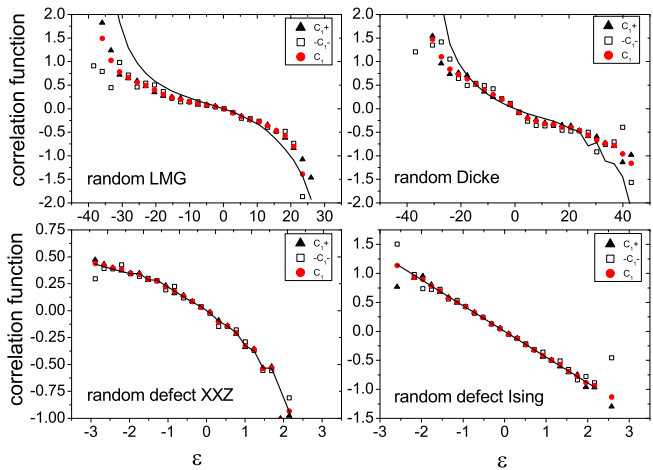}
\caption{ (Color online) Similar to Fig.\ref{fig-C1}, but for
 the correlation functions  ${\cal{C}}_1^{\pm}$ in Eq.(\ref{C1pm}) and $\ww {\cal C}_1$ in Eq.(\ref{wwC1})
  in modified versions of the four models,
 in which $30\%$ of the nonzero $V_{ij}$ are negative.
 The solid curves are given by the theoretical prediction on the right hand side of Eq.(\ref{C1ran}).
 }\label{cfrandom}
\end{figure}

 \section{Correlation functions for $V_{ij}$ with nonhomogeneous signs}\label{sect-cf-randomsign}

 In this section, we discuss the case that nonzero elements $V_{ij}$ have both positive
 and negative signs.
 In this case, nonzero $V_{ij}$ have quite strong fluctuations, such that Eq.(\ref{C1-expr}) does not hold.

 We find that sign-correlation still exists among $ C_{\alpha i}$ and $C_{\alpha j}$, for
 those unperturbed states that are coupled by the perturbation $V$.
 To see this point, let us divide the set $S_1$ into two subsets
 according to the sign of $V_{ij}$, denoted by $S_1^{\pm}$, respectively.
 We use ${\cal{C}}_1^{\pm}$ to denote the corresponding (first-order) correlation functions,
 defined by
\begin{equation}\label{C1pm}
 {\cal{C}}_1^{\pm} = \la C_{\alpha i}C_{\alpha j}\ra^{\pm}/\Pi(\varepsilon)
 \quad \text{for} \ (i,j) \in S_1^{\pm}.
\end{equation}
 Following arguments similar to those leading to Eq.(\ref{C1-expr}), one gets
\begin{equation}\label{c1-rela-1}
 {\ov V_{+} \ov N_+}{\cal{C}}_1^{+} + {\ov V_{-} \ov N_-}
{\cal{C}}_1^{-} \simeq -{\varepsilon},
\end{equation}
 where $\ov V_{\pm}$ and $\ov N_{\pm}$ are defined in a way similar to $\ov V$ and $\ov N$
 discussed previously, but with respect to the sets $S_1^{\pm}$, respectively.

 Let us define a correlation function weighted by the sign of $V_{ij}$, denoted by $\ww {\cal C}_1$,
\begin{gather}\label{wwC1}
 \ww {\cal C}_1= \la \text{sign} (V_{ij})
 C_{\alpha i}C_{\alpha j}\ra  /\Pi(\varepsilon), \quad  (i,j) \in S_1 ,
\end{gather}
Similar to Eq.(\ref{C1-expr}), it is found that
\begin{equation}\label{C1ran}
 \ww{\cal C}_1(\varepsilon) \simeq -{\varepsilon}\left( {\ov{|V|} \ov N} \right)^{-1},
\end{equation}
 where $\ov{|V|} = \la |V_{ij}|\ra$.

 For the simplicity in discussion, let us consider the specific case that $\ov V_{+} =
 -\ov V_{-} =\ov{|V|}$.
 Then, Eqs.(\ref{C1ran}) and (\ref{c1-rela-1}) give
\begin{equation}\label{}
 \ov N \ww {\cal C}_1 \simeq \ov N_+ {\cal{C}}_1^{+}
 - N_- {\cal{C}}_1^{-}.
\end{equation}

 Noting that $\ov N_+ + \ov N_- = \ov N$,
 it would be reasonable to expect that
\begin{equation}\label{3C}
  \ww {\cal C}_1 \simeq  {\cal{C}}_1^{+} \simeq -{\cal{C}}_1^{-}.
\end{equation}
 This suggests that, for pairs $(i,j)$ in the set $S_1$,
 $C_{\alpha i}C_{\alpha j}$ tend to have the same sign as $(-\varepsilon \text{sign}(V_{ij}))$.
 Note that this phenomenon has also been observed in the homogeneous-sign case discussed
 in the previous section [see Eq.(\ref{C1-expr})].

\begin{figure}
\includegraphics[width=\columnwidth]{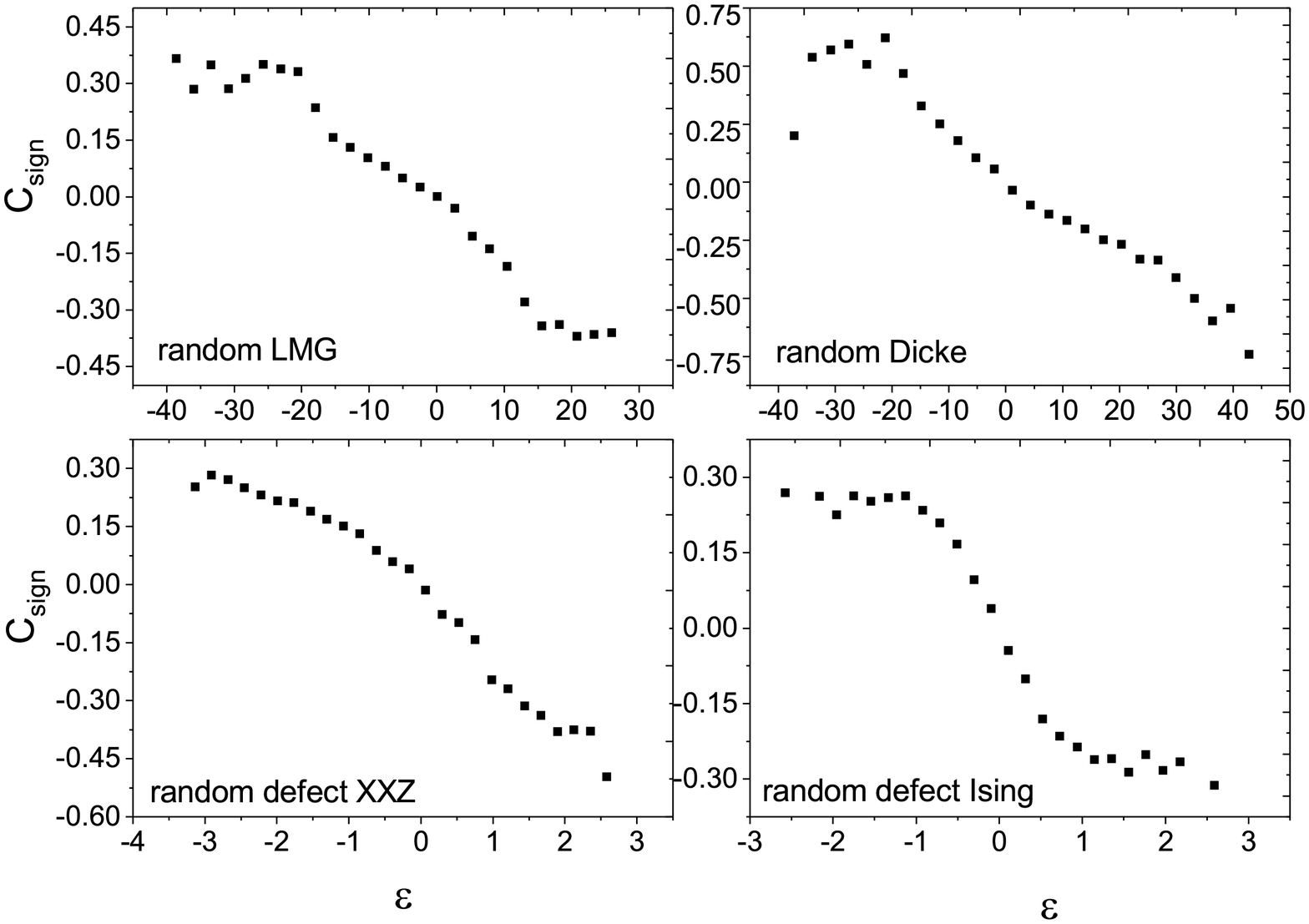}
\caption{ Sign correlation between directly coupled components of the EFs,
 ${\cal C}_{\rm sign}$ defined in Eq.(\ref{cfs}).
 }\label{cfsign}
\end{figure}

 To test whether the expectation in Eq.(\ref{3C}) is correct, we have studied modified versions
 of the four models discussed above,
 changing the signs of a percentage of randomly-chosen nonzero elements $V_{ij}$ to the negative one.
 For brevity, we call the models thus obtained the random LMG model, and so on.
 Our numerical simulations confirm the validity of Eq.(\ref{3C}) in all the four
 modified models and show that Eq.(\ref{C1ran}) works well except in the
 tail regions of the EFs in the LMG and the Dicke models (Fig.\ref{cfrandom}) \cite{foot-C1-}.
 The sign correlation between $C_{\alpha i}$ and $C_{\alpha j}$
 has been observed in a direct computation of the following quantity,
\be \label{cfs}
{\cal{C}}_{\rm sign}= \left \langle {\rm sign}(C_{\alpha i}C_{\alpha j}) \cdot {\rm sign}(V_{ij})
 \right \rangle_{\varepsilon} \quad \text{for} \ (i,j)\in S_1.
\ee
 As seen in Fig.\ref{cfsign},  the sign correlation increases with increasing $|\varepsilon|$.

 \section{An application}\label{sect-apply}
 As an application of the above results, let us consider the transition
 probability from an initial state $|E^0_i\ra$ to final states $|E^0_j\ra$ with direct
 coupling ($V_{ij}\ne 0$), which we denote by $F_i(t)$,
\begin{equation}\label{Fi-t}
 F_i(t) = {\sum_{j\in g_i}} |F_{ij}(t)|^2.
\end{equation}
 For simplicity in discussion, we consider cases satisfying the following requirements:
 nonzero elements $V_{ij}$ are close to each other,  the
 values of $|C_{\alpha j}|$ do not have large fluctuations with respect to the label $j$,
 and the $\varepsilon$-dependence of $\ov V$ and $\ov N$ can be neglected.

\begin{figure}
\includegraphics[width=\columnwidth]{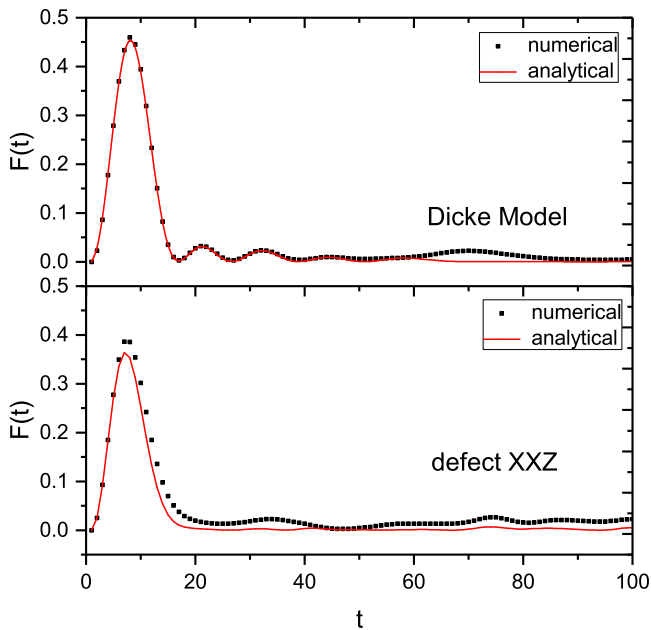}
\caption{ (Color online) Variation of the transition probability $F_i$ in Eq.(\ref{Fi-t})
 (solid squares).
 Solid curves indicate predictions of Eq.(\ref{Fi-final}).
 The times are given in unit of $\tau=10^{-3}/d $,
 where $d$ is the averaged level spacing.
 }\label{fig-ft}
\end{figure}

 Let us first discuss variation of $F_{ij}$ with $j$.
 To this end, using $E_\alpha=E^0_i-\varepsilon_{\alpha i} $, we write it as
\begin{equation}\label{}
 F_{ij} = \left( \sum_\alpha C_{\alpha i} C_{\alpha j}e^{i\varepsilon_{\alpha i}t} \right)
 e^{-iE^0_it}.
\end{equation}
 According to Eq.(\ref{C1-expr}),
 $C_{\alpha i}C_{\alpha j}$ are on average proportional to $-\varepsilon \Pi(\varepsilon)$,
 hence, the main contribution to $F_{ij}$ should come from
 those perturbed states $|E_\alpha\ra$ for which
 $|\varepsilon_{\alpha i}\Pi(\varepsilon_{\alpha i})|$ are large.
 For these perturbed states, as discussed previously,
 $C_{\alpha i}C_{\alpha j}$ tend to have the same sign as
 $(-\varepsilon_{\alpha i}V_{ij})$.
 Noting the homogeneousness of the sign of nonzero $V_{ij}$ and the
 smallness of the fluctuation of $|C_{\alpha j}|$ with $j$,
 it is seen that, on average, $F_{ij}$ do not have large fluctuation with $j$ for $j\in g_i$.

\begin{figure}
\includegraphics[width=\columnwidth]{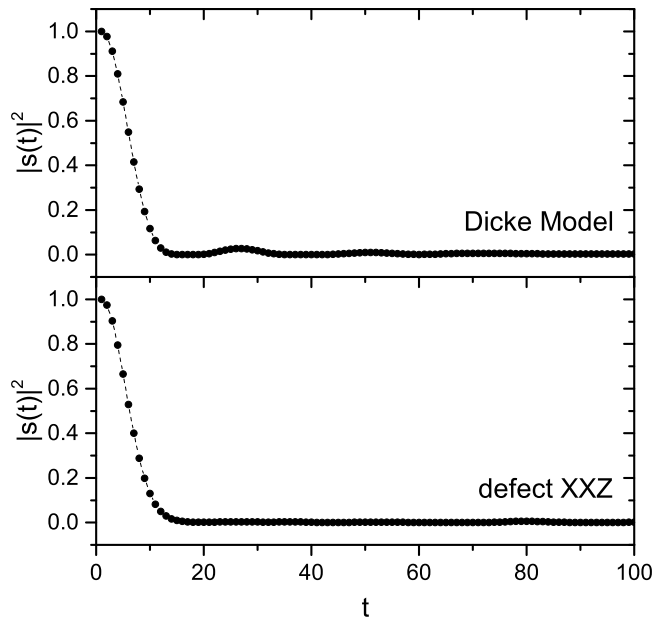}
\caption{  Variations of the survival probability $|s(t)|^2$ under the same initial conditions
  as those for Fig.{\ref{fig-ft}}.
 }\label{fig-st}
\end{figure}

 Then, we get the following approximation for these labels $j$,
 \be
 F_{ij}(t)\simeq \frac{1}{ \ov N}\sum_{j'\in g_i}F_{ij'},
 \ee
 and, as a result, the following expression of  $F_i(t)$,
\be \label{ft2}
 F_i(t) \simeq \ov N \left| \frac{1}{ \ov N}\sum_{j'\in g_i}F_{ij} \right|^2=
  \frac{1}{ \ov N} \left|\sum_{\alpha}\sum_{j\in g_i}
 C_{\alpha i} C_{\alpha j} e^{i\varepsilon_{\alpha i}t}\right|^2,
\ee
 Further, due to the assumed small fluctuation of nonzero $V_{ij}$, one has
 $F_i(t) \simeq \frac{1}{ \ov N} \left|\frac{1}{\ov V }\sum_{\alpha}\sum_{j\in g_i}V_{ij}C_{\alpha j}
 C_{\alpha i}e^{i\varepsilon_{\alpha i}t}\right|^2 \nonumber $.
 Finally, making use of Eq.(\ref{Seq}), one gets the following expression,
\begin{gather} \notag
 F_i(t) \simeq \frac{1}{ \ov N} \left|\frac{1}{\ov V }\sum_{\alpha}
 \varepsilon_{\alpha i}|C_{\alpha i}|^2 e^{i\varepsilon_{\alpha i}t}\right|^2 
 \\ \simeq \frac{1}{ \ov N (\ov V)^2} \left| \frac{\partial s_i (t)}{\partial t}  \right|^2, \label{Fi-final}
\end{gather}
where
\be
 s_i(t)= \la E^0_i|e^{-i(H-E^0_i)t}|E^0_i\ra.
\ee
 It is easy to see that,
 apart from a phase factor, $s_i(t)$ gives the survival probability amplitude for the initial state.

 To test numerically the prediction of Eq.(\ref{Fi-final}), we consider the Dicke model
 and the defect XXZ model.
 (The LMG model and the defect Ising model do not meet the requirements discussed above.)
 Numerical simulations show that Eq.(\ref{Fi-final}) works well in the Dicke model and
 works approximately in the defect XXZ model
 (see Fig.\ref{fig-ft} for two examples).
 Examples for the survival probabilities in these two models are shown in Fig.{\ref{fig-st}}

\section{Conclusions}\label{sect-sum}

 In summary, in this paper, we have studied correlation functions with respect to the energy difference
 between perturbed and unperturbed states, in quantum chaotic systems
 whose Hamiltonian matrices have a sparse structure in the unperturbed bases.
 Analytical expressions have been derived for some correlation functions
 and have been tested in numerical simulations performed in four models.
 An application is given to a property of a transition probability.
 It should be reasonable to expect that more applications may be found in future investigations.

 \acknowledgements

 The authors are grateful to J.Gong for valuable discussions.
 This work was partially supported by the Natural Science Foundation of China under Grant
 Nos.~11275179 and 11535011,
 and the National Key Basic Research Program of China under Grant
 No.~2013CB921800.


\begin{thebibliography}{99}
\bibitem{foot-Vii} In the case that some of $V_{ii}$ are nonzero, one may move
 $\sum_i |E_i^0\ra V_{ii} \la E^0_i|$ to the unperturbed Hamiltonian, with the total Hamiltonian unchanged.
\bibitem{foot-C1-} Large fluctuations of ${\cal C}_1^{-}$
at the largest and smallest values of $\varepsilon$  in Fig.\ref{cfrandom}
 are due to smallness of the number of the items, which are available for taking average.
\end{thebibliography}
\end{document}